\documentclass[journal, 10pt]{IEEEtran}
\IEEEoverridecommandlockouts
\usepackage[noadjust]{cite}
\usepackage{amsmath,amssymb,amsfonts}
\usepackage{algorithm}
\usepackage{algorithmic}
\usepackage{graphicx}
\usepackage{textcomp}
\usepackage{xcolor}
\usepackage{tikz}
\usepackage{flushend}

\usetikzlibrary{shapes,arrows,shadows,positioning}
\usetikzlibrary{arrows, arrows.meta, shapes, shapes.multipart, trees, positioning,
	backgrounds, fit, matrix, petri, calc, automata, chains}
\usepackage{ellipsis}
\usetikzlibrary{calc}
\usetikzlibrary{decorations.pathreplacing,decorations.markings,shapes.geometric}
\usetikzlibrary{patterns}
\def\BibTeX{{\rm B\kern-.05em{\sc i\kern-.025em b}\kern-.08em
    T\kern-.1667em\lower.7ex\hbox{E}\kern-.125emX}}

\author{
	\IEEEauthorblockN{Onur Ayan, H. Murat G\"ursu, Arled Papa, Wolfgang Kellerer}\\
	\IEEEauthorblockA{Chair of Communication Networks}\\
	\IEEEauthorblockA{Technical University of Munich, Germany}\\
	\IEEEauthorblockA{Email: \{onur.ayan, murat.guersu, arled.papa, wolfgang.kellerer\}@tum.de}
}
\title{Probability Analysis of Age of Information in Multi-hop Networks}

\newcommand{\E}{\mathbb{E}}

\definecolor{myred}{RGB}{220,43,25}
\definecolor{mygreen}{RGB}{0,146,64}
\definecolor{myblue}{RGB}{0,143,224}
\definecolor{mygray}{gray}{0.90}

\usepackage[absolute,showboxes]{textpos}

\TPMargin{5pt}

\newcommand{\copyrightstatement}{
	\begin{textblock}{0.84}(0.08,0.01)    
		\noindent
		\footnotesize
		\copyright 2020 IEEE. Personal use of this material is permitted. Permission from IEEE must be obtained for all other uses, in any current or future media, including reprinting/republishing this material for advertising or promotional purposes, creating new collective works, for resale or redistribution to servers or lists, or reuse of any copyrighted component of this work in other works.
	\end{textblock}
}

\begin{document}
	
\copyrightstatement

\maketitle

\begin{abstract}
Age-of-information (AoI) is a metric quantifying information freshness at the receiver. It captures the delay together with packet loss and packet generation rate. However, the existing literature focuses on average or peak AoI and neglects the complete distribution. In this work, we consider a $N$-hop network with time-invariant packet loss probabilities on each link. We derive closed form equations for the probability mass function of AoI. We verify our findings with simulations. Our results show that the performance indicators considered in the literature such as average or peak AoI may give misleading insights into the real AoI performance.
\end{abstract}

\section{Introduction}
\textit{Age of information} (\textit{AoI}) is a metric that measures the information freshness from the perspective of the receiver monitoring a remote process. It is defined as the elapsed time since the generation of the freshest packet at the receiver \cite{kaul2012real}. AoI increases until the arrival of a fresher status update at the receiver and drops upon its successful reception. Therefore, receiving status updates  regularly plays a key role in sustaining information freshness. In addition, the staleness of a newly received update is characterized by the time it has spent in the network to reach the destination. As a result, a good AoI performance is achieved when status updates are delivered not only regularly but also timely.

While AoI is applicable to almost any cyber-physical system scenario, some of the most prominent applications are vehicular networks \cite{kaul2011minimizing}, unmanned aerial vehicles \cite{liu2018age, abdelmagid2019average} and networked control systems \cite{ayan2019age, champati2019performance} where periodic status updates are being sent over a wireless network. In such a setting, packets may need to traverse multiple hops towards the destination where each link is prone to delay and packet loss. Consequently, selecting a different path may result in higher or lower AoI at the receiver, where monitoring and decision making mechanism resides. Hence, utilization of outdated information for decision making may lead to performance degradation of the application and even damage the physical environment. Therefore, obtaining detailed analytical models of the AoI becomes crucial to gain insights into the expected performance of underlying applications. 

 
Vast majority of previous work proposes optimal scheduling \cite{hsu2017age, kadota2016minimizing, kadota2018optimizing, he2018optimal, yates2017status} or queuing policies \cite{kaul2012status, costa2016on, sun2017update} to minimize the expected AoI in single-hop wireless networks. On the other hand, \cite{he2016on, abdelmagid2019average} focus on peak age metric which considers the AoI only at instances of a new update. Hence, the peak AoI is a measure for the ``worst case scenario'' but it does not provide information about the real AoI performance. To the best of our knowledge, \cite{inoue2019general}, which considers a single-hop scenario, is the only work addressing the probability distribution of AoI. In particular, the authors derive a general formula for the stationary distribution of AoI for single-server queuing systems which is applicable to different service disciplines.

Moreover, in \cite{kaul2012status}, authors show that, AoI can be decreased if a more recent information always replaces an older one in the transmission queue. This insight is extended to a multi-hop scenario in \cite{bedewy2017age}. In \cite{bedewy2017age}, authors assume exponentially distributed transmission times over the links and show that the preemptive Last Generated First Served (PLGFS) queuing policy is age-optimal. Furthermore, the average AoI for PLGFS queues is addressed in \cite{yates2018age} which considers a multi-hop line network and characterizes the average AoI. 

One of the most related work is \cite{farazi2019fundamental} which considers a multi-source, multi-monitor scenario in a multi-hop setting. The authors derive lower bounds on the instantaneous peak and average AoI by employing fundamental graph-theoretical measures such as connected domination number and average shortest path length.

The work closest to ours is \cite{talak2017minimizing} which proposes an optimal stationary scheduling policy to minimize AoI in a lossy multi-hop line network. They show that the  optimal policy for the multi-hop problem can be obtained by solving an equivalent problem separately on each link over the path. They derive a closed form expression for the expected AoI at the receiver given the activation frequency of each link.

Concluding on the previous work, we observe that the existing literature on multi-hop networks is limited to assessment of the performance metrics such as peak and expected AoI. However, the peak AoI might be inconclusive and the expected AoI is not sufficient when it comes to performance guarantees for real-time, physical systems. Moreover, the same expected AoI can stem from two distinctive distributions which may result in different system performances. Contrarily, the probability distribution of AoI can provide the exact insights into AoI and thus age dependent performance.

To the best of our knowledge, the probability distribution of AoI in a multi-hop network is not yet covered in the existing literature. We close this gap and provide the probability mass function (PMF) of any AoI $\Delta$ as a function of stationary loss probabilities on each link between source and destination that are $N$-hop away from each other. We model the AoI as joint probabilities of the links over the path and show that occurrence probability of any AoI can be calculated in a scalable manner recursively from the $(n$-$1)$-th hop.


\subsection{Notations} 
Throughout this paper $\E[X]$ stands for the expected value of a random variable $X$. $\text{Pr}[A ~|~ B] $ denotes the conditional probability, i.e., the probability of $A$ given $B$. The set of positive integers is denoted by $\mathbb{Z}^+$. 
\section{System Model}
\label{sec:systemmodel}

We consider a physical process located at the source node generating status updates periodically and a monitor located at the receiver node. The source and receiver nodes are $N$-hops away from each other. We assume a one-dimensional multi-hop topology, which is also called a line network in the literature. Note that, once the routing takes place, one can treat the established path between any source-destination pair as a line network. 

Each transmitter over the path, i.e., the source and the intermediate relay nodes, discards any older packet in the transmission queue upon the arrival of a new update\footnote{Under the assumption that status is Markovian, having  received an  update, the  receiver does not benefit from the reception of older status updates. Thus, older packets are considered to be obsolete and ``non-informative''.}. As a result, there is no queuing effect in our model as there is always a single packet to be forwarded. A new update at any node is re-transmitted until it is successfully received by its next hop or replaced with a more recent information.

As in a typical multi-hop scenario, we assume the nodes to be spatially distributed. Therefore, each link $n$ is prone to packet loss with time-invariant failure probability, i.e., $p_n(t) = p_n, \, \forall t$. 
We assume Rayleigh block fading model to represent the wireless medium behavior such that the average failure probability of each packet is independent. As a result, the outcome of a transmission on the $n$-th link can be abstracted as a Bernoulli trial with a constant failure probability $p_n$.

\begin{figure}[t]
	\centering
	\resizebox{\columnwidth}{!}{\begin{tikzpicture}
		
\foreach \x in {0,...,7} 
	\foreach \y in {-1,...,2}
		\filldraw[fill=white, draw=black] (\x, \y) rectangle (\x + 1, \y + 1);

%



\draw[pattern=north east lines, pattern color=orange] (0.5,2.5) circle (10pt) node[align=left] (sourceGreen) {$\mathbf{k}$};
\draw[pattern=north east lines, pattern color=orange] (1.5,1.5) circle (10pt) node[align=left] (relay1Green) {$\mathbf{k}$};
\draw[pattern=north east lines, pattern color=orange] (2.5,0.5) circle (10pt) node[align=left] (relay2Green) {$\mathbf{k}$};
\draw[pattern=north east lines, pattern color=orange] (3.5,-0.5) circle (10pt) node[align=left] (receiverGreen) {$\mathbf{k}$};
\draw[pattern=north east lines, pattern color=orange] (7.5,2.5) circle (10pt) node[align=left] () {{$\mathbf{k}$+$\mathbf{1}$}};

\draw[->, dashed, black, thick] (sourceGreen.south east)  to[bend left] (relay1Green.north west);
\draw[->, dashed, black, thick] (relay1Green.south east)  to[bend right] (relay2Green.north west);
\draw[->, dashed, black, thick] (relay2Green.south east)  to[bend left] (receiverGreen.north west);

\newcommand{\taxisY}{-2}
\node[left, black, align=center] (t) at (-1, \taxisY){Process\\timeline};

\node[](tAxis) at (9, \taxisY) {};
\coordinate[] (torigin) at (-1, \taxisY);
\draw[->, dashed] (torigin) -- (tAxis);

\draw[black,fill=black] (0,\taxisY) circle (1.5pt);
\draw[black,fill=black] (7,\taxisY) circle (1.5pt);

\node[below] (t) at (0, \taxisY){$k$};
\node[below] (t-1) at (7, \taxisY){$k+1$};
\draw [<->, thick] (0, -2.5) -- node[align=left, below] {$t_p$} (7, -2.5);

\draw [<->, thick] (1, -1.25) -- node[align=left, below] {$t_s$} (2, -1.25);

\newcommand{\caxisY}{-3}

\draw[black,fill=orange] (0,\taxisY) circle (1.5pt);
\draw[->, thick, orange] (0, \taxisY) -- (0,\taxisY+.55);
\draw[black,fill=orange] (7,\taxisY) circle (1.5pt);
\draw[->, thick, orange] (7, \taxisY) -- (7,\taxisY+.55);

\draw [decorate,decoration={brace,amplitude=4pt,mirror}] (-0.1,2.9) -- (-0.1,2.1) 
node [black,midway, xshift=-0.5cm,align=left] {\scriptsize Source };
\draw [decorate,decoration={brace,amplitude=4pt,mirror}] (-0.1,1.9) -- (-0.1,1.1) 
node [black,midway, xshift=-0.5cm,align=left] {\scriptsize Relay 1 };
\draw [decorate,decoration={brace,amplitude=4pt,mirror}] (-0.1,0.9) -- (-0.1,0.1) 
node [black,midway, xshift=-0.5cm,align=left] {\scriptsize Relay 2 };
\draw [decorate,decoration={brace,amplitude=4pt,mirror}] (-0.1,-0.1) -- (-0.1,-.9) 
node [black,midway, xshift=-0.5cm,align=left] {\scriptsize Dest. };

\end{tikzpicture}}
	\caption{Example 3-hop scenario, i.e., $N = 3$ with a sampling period of 7 slots, i.e., $m = 7$. Together with dashed lines, the orange circle illustrates the path of the $k$-th update. The next status update, $k+1$, is available $m = 7$ slots after the previous sampling event.
	}
\label{fig:timemodel}
\end{figure}
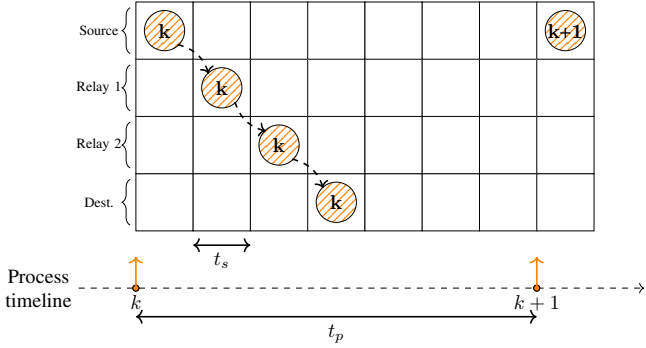

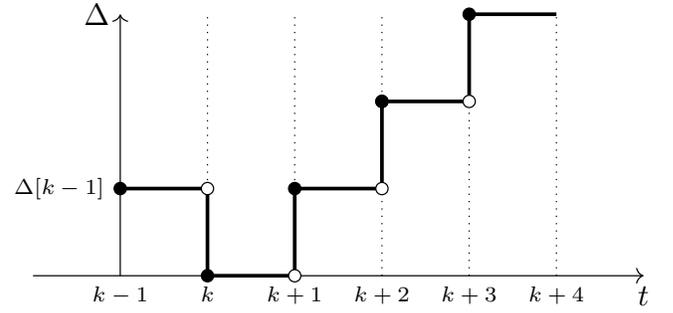
\begin{figure}[t]
	\centering
	\resizebox{\columnwidth}{!}{\begin{tikzpicture}

\draw[->] (-1,0) -- (6,0) node[anchor=north] {$t$};
\draw	(0,0) node[anchor=north] {\scriptsize$k-1$}
		(1,0) node[anchor=north] {\scriptsize$k$}
		(2,0) node[anchor=north] {\scriptsize$k+1$}
		(3,0) node[anchor=north] {\scriptsize$k+2$}
		(4,0) node[anchor=north] {\scriptsize$k+3$}
		(5,0) node[anchor=north] {\scriptsize$k+4$};

\draw[->] (0,0) -- (0,3) node[anchor=east] {$\Delta$};
\draw[dotted] (1,0) -- (1,3);
\draw[dotted] (2,0) -- (2,3);
\draw[dotted] (3,0) -- (3,3);
\draw[dotted] (4,0) -- (4,3);
\draw[dotted] (5,0) -- (5,3);

\draw[very thick, black, solid] (0,1) -- (1,1) -- (1,0) -- (2,0) -- (2,1) --  (3,1) -- (3,2) -- (4,2) -- (4,3) -- (5,3);

\draw (-.7, 1) node {\scriptsize$\Delta[k-1]$}; 
\node[circle,draw=black, fill=white, inner sep=0pt,minimum size=4pt] (b) at (1,1) {};
\node[circle,draw=black, fill=white, inner sep=0pt,minimum size=4pt] (b) at (2,0) {};
\node[circle,draw=black, fill=white, inner sep=0pt,minimum size=4pt] (b) at (3,1) {};
\node[circle,draw=black, fill=white, inner sep=0pt,minimum size=4pt] (b) at (3,2) {};
\node[circle,draw=black, fill=white, inner sep=0pt,minimum size=4pt] (b) at (4,2) {};

\node[circle,draw=black, fill=black, inner sep=0pt,minimum size=4pt] (b) at (0,1) {};
\node[circle,draw=black, fill=black, inner sep=0pt,minimum size=4pt] (b) at (1,0) {};
\node[circle,draw=black, fill=black, inner sep=0pt,minimum size=4pt] (b) at (2,1) {};
\node[circle,draw=black, fill=black, inner sep=0pt,minimum size=4pt] (b) at (3,2) {};
\node[circle,draw=black, fill=black, inner sep=0pt,minimum size=4pt] (b) at (4,3) {};
\end{tikzpicture}}
	\caption{An example evolution of the AoI at the receiver. A new status update is received in sampling period $k$. During the following 3 sampling periods, the monitor fails to update its most recent information.
	}
	\label{fig:ageplot}
\end{figure}

Time is divided into slots of length $t_s$ which is also the smallest time unit in our model. Each packet transmission starts at the beginning of a slot and completes within the same slot. The generation of status updates is periodic with $t_p$ slots between two consecutive updates, i.e., $m \cdot t_s = t_p,$ with $m \in \mathbb{Z}^{+}$. We call the generation of an update packet a \textit{sampling event} and the time between two consecutive sampling events a \textit{sampling period}\footnote{Periodic sampling is a well-established way of generating status updates in sensor and actuator networks. In most of the cases, sampling periods are much longer than network time slots, i.e., $m \gg 1$.}.

We assume that transmission schedule of intermediate nodes within a sampling period is designed in such a way that each link is activated in the same order as they appear over the path. Fig.~\ref{fig:timemodel} depicts an example of such a schedule where the Source-to-Relay 1 link is activated in the first slot. The intermediate nodes are scheduled $2$ and $3$ slots after the sampling event. 
Similar transmission schedules for multi-hop networks have been suggested in \cite{alur2009modeling} and enable the reception of a newly generated update within the same sampling period after $N$-hops. Note that, this implies $m \geq N$. In the following analysis, we assume that this inequality holds.


Motivated by the \textit{discrete-time} control systems \cite{astrom2008feedback}, in which the model evolves in discrete steps in time and the system status is considered unchanged until the next sampling event, we assume the dynamics of the AoI to be discrete as well. Therefore, we evaluate the AoI at the receiver only once in each sampling period and right before the next sampling event. Fig. \ref{fig:ageplot} depicts an example time evolution of the AoI at the receiver when a new status update is received in sampling period $k$. Note that, while the AoI follows a ``staircase'' model in continuous time, from the discrete-time system's perspective it is a linear increase as we evaluate the AoI only once and at the end of each sampling period.  
This is similar to the approach in \cite{kadota2016minimizing} which only allows the AoI to decrease at the end of each sampling period. 

Let $\gamma_n[k] \in \{0, \, 1\}$, $n \in \{1, \, 2, \, \dots, \, N\}$, indicate the outcome of the transmission on the $n$-th link in sampling period $k$ with $\text{Pr}[\gamma_n[k] = 1 ] = 1 - p_n, \, \forall k$. Furthermore, let $\Delta_n[k]$, $n \in \{0,\, 1, \, 2, \, \dots, \, N\}$, be the AoI at hop $n$ \textit{after the transmission slot that is allocated to the the previous link}. As a result, the discrete time model of the AoI follows as:

\begin{equation}
\Delta_{n}[k] =
\begin{cases}
\Delta_{n-1}[k] &, \gamma_n[k] = 1\\
\Delta_{n}[k-1] + 1 &, \gamma_n[k] = 0\\
\end{cases}
\label{eq:age_model}
\end{equation}
with $\Delta_0[k] = 0, \, \forall k$. Note that $\Delta_0[k]$ denotes the AoI at the source node and it always equals to zero due to packet discarding policy. $\Delta_{N}[k]$ denotes the AoI at the monitor at the end of the $k$-th sampling period.

Our model considers that each hop has a single slot between two sampling events represented by a single loss probability. Allocation of multiple slots in a consecutive fashion can be incorporated into the model through merging the loss probabilities. Let $p_{l}^*$ denote the loss probability of a single transmission on link $l$ and $L$ the number of consecutive slots allocated to $l$. In such a setting, we can group them together and determine the loss probability as $p_l = (p^*_l)^L$. The model remains unchanged as long as the slots are allocated in a consecutive fashion and the total number of used slots does not exceed the sampling period. 
\section{Analysis}
\label{sec:analysis}

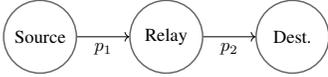
\begin{figure}[t]
	\centering
	\resizebox{.5\columnwidth}{!}{
		\begin{tikzpicture}[
		circ/.style={circle, draw=black!60, thick,minimum width=1.4cm, minimum height= 10mm},
		]
		\node[circ]      (source)                              {Source};
		\node[circ]      (relay)            [right=of source]                  {Relay};
		\node[circ]      (destination)       [right=of relay] {Dest.};
		\draw[->] (source.east)-- node[anchor=north] {$p_1$} (relay.west);
		\draw[->] (relay.east)-- node[anchor=north] {$p_2$} (destination.west);
		\end{tikzpicture}}
	
	\caption{Packet is sent from source to the relay and forwarded to the destination when there is no direct link.}
	\label{2_hop}
\end{figure}

We begin with the simple single-hop scenario as illustrated from Source to Relay in Fig. \ref{2_hop}. 
As the packet is always fresh at the source, in case of a success the age at the relay is reset to zero.
The failure  probability on the first link is denoted with $p_1 \in [0, 1]$. Thus, the probability of the AoI at the relay being $\delta_1 \in \mathbb{Z}_{\geq 0} $ can be written as:
\begin{equation}
\text{Pr}\left[\Delta_1[k] = \delta_1 \right] = (1-p_1) \cdot p_1^{\delta_1}, \quad \forall k.
\label{eq:1_hop_age}
\end{equation}
Eq.~\eqref{eq:1_hop_age} can be interpreted as the probability of $\delta_1$ unsuccessful transmission attempts following a successful transmission. 
Thus, the expected AoI at hop $1$ follows as:
\begin{equation}
\mathbb{E}[\Delta_1] = \sum_{\delta_1 = 0}^{\infty}{\text{Pr}\left[\Delta_1[k]=\delta_1\right] \cdot \delta_1} = \frac{p_1}{1-p_1}. 
\label{eq:expected1}
\end{equation}

Consider a 2-hop scenario as in Fig.~\ref{2_hop} with constant loss probabilities $p_1$ and $p_2$ on the source-to-relay and the relay-to-destination links, respectively. Given the loss probabilities on two links are independent, we can treat each link independently.

The contribution of the source-to-relay link to $\Delta_1$, is analogous to the single-hop case. However, in contrast to the source node, the information at the relay is now $\delta_1$ periods old. Let us denote the sampling period, in which the freshest information is received by the $n$-th hop as $k_n$. Since the AoI at the previous hop and at the $n$-th hop are equal at that instance, we can write $\Delta_{n-1}[k_n] = \delta_{n-1}$.
Consequently, using this equality we can treat the age at the second hop as a further aging process, i.e., given $\Delta_1[k_2] = \delta_1$, the probability to have $\Delta_2[k] = \delta_2$ after the second link can be calculated as:

\begin{flalign}
\quad \text{Pr}[\Delta_2[k] &= \delta_2 ~|~ \Delta_1[k_2]=\delta_1]  \nonumber && \\ 
&=
\begin{cases}
0  &\text{ if }  \delta_2 < \delta_1 \\
(1-p_2) {p_2}^{\delta_2 - \delta_1}  &\text{ if }  \delta_2 \geq \delta_1
\end{cases}.
\label{eq:marg_2hop}
\end{flalign}

Here, we exploit the line-network topology and the strictly increasing property of AoI. That is, the age at the destination cannot be lower than the age at the first hop since there is no other path between source-destination pair. This results in zero probability for all ages below $\delta_1$.  

$\delta_1$ represents the first link’s contribution to the total age $\Delta_2 [k] = \delta_2$. Furthermore, $\delta_2 - \delta_1$ is the elapsed time since the most recent information was received by the destination and can also be interpreted as the second link's contribution to the total age $\delta_2$. 
As a result, we can use the law of total probability to formulate the probability of $\Delta_{2}$ being $\delta_2$ as:
\begin{align}
\text{Pr}\left[\Delta_2[k]=\delta_2\right]\nonumber = \sum_{\delta_1=0}^{\delta_2} & \text{Pr}\left[\Delta_2[k]=\delta_2 ~|~ \Delta_1[k_2]=\delta_1\right]\\  \cdot& \text{Pr}\left[\Delta_1[k_2]=\delta_1\right], \quad \forall k.
\label{eq:2_hop_age_1}
\end{align}
Note that we were able to merge the two cases from Eq. \eqref{eq:marg_2hop} as we are concerned only with the nonzero summands. By plugging Eq.~\eqref{eq:1_hop_age} and Eq.~\eqref{eq:marg_2hop} in we get:
\begin{equation}
\text{Pr}\left[\Delta_2[k]=\delta_2\right]= (1-p_1)(1-p_2)\cdot \frac{{p_2}^{\delta_2+1}-{p_1}^{\delta_2+1}}{{p_2}-{p_1}}.
\label{eq:2_hop_age_4}
\end{equation}
%
The proof of Eq.~\eqref{eq:2_hop_age_4} and the following equations are given in appendix \ref{sec:appendic}. For $p_1 = p_2$, one can use the first line of Eq.~\eqref{eq:appendix2hop}.

Analogous to Eq.~\eqref{eq:expected1}, it can be shown, that the expected AoI at hop $2$ is:
\begin{equation}
\E[\Delta_2] = \frac{p_1}{1 - p_1} + \frac{p_2}{1 - p_2}.
\end{equation}
In fact, the expected AoI at any hop $n$ can be obtained from:
\begin{equation}
\label{eq:3hopexpectedAoI}
\E\left[\Delta_{n}\right] = \sum_{i=1}^{n} \frac{p_i}{1-p_i}.
\end{equation}


%
%
%
%
Next, we extend our results to $n$-hop, 
\begin{align}
\text{Pr}\left[ \Delta_n[k]=\delta_{n} \right] = \sum_{\delta_{n-1}=0}^{\delta_{n}}&  \text{Pr}\left[ \Delta_{n}[k] = \delta_{n} ~|~ \Delta_{n-1}[k_n] = \delta_{n-1}\right] \nonumber \\ \cdot &\text{Pr}\left[\Delta_{n-1}[k_n] = \delta_{n-1}\right].
\label{eq:n_hop_age_1}
\end{align}
Closed form for higher number of hops can also be obtained using a similar analysis. For instance for 3-hops we obtain:
%

\begin{align}
\label{eq:cf3hop}
&~ \text{Pr}\left[\Delta_3[k] = \delta_{3}\right]  \nonumber && \\
&= \frac{\prod^3_{i=1} (1-p_i)}{p_2-p_1} \cdot \sum_{j=1}^2 (-1)^j \cdot {p_j} \cdot \frac{{p_3}^{\delta_3+1}-{p_j}^{\delta_3+1}}{{p_3}-{p_j}} .
\end{align}

In Alg.~\ref{alg:recursive} a pseudo-code for a recursive calculation of probabilities for any $n$ hop is given. The algorithm has a time complexity of $O((\delta_n+1) \cdot n)$.
\begin{algorithm}[t!]
\caption{Recursive age function: $f(\delta_n,n,\mathbf{p}^n) = o$}
\label{alg:recursive}
\begin{algorithmic}
\item[\textbf{Input:}] $\delta_n$ age, $n$ number of hops, $\mathbf{p}^n$ vector of loss probabilities for n hops
\item[\textbf{Output:}] $o$ the probability of age $\delta_n$ with n hops
\STATE Initialize: $o\leftarrow0$
\IF{$n = 1$}
	\STATE \textbf{return} $(1-p_n) \cdot {p_n}^{\delta_n}$
\ELSE
	\FOR{$\delta_{n-1},\,\in [0 \,\, \delta_n]$}
		\STATE $o \leftarrow o + \left((1-p_n) \cdot {p_n}^{\delta_n - \delta_{n-1}}\right)\cdot f(\delta_{n-1},{n-1},\mathbf{p}^{n-1})$
	\ENDFOR
	\STATE \textbf{return} $o$
\ENDIF
\end{algorithmic}
\end{algorithm}



\begin{figure}[t]
	\centering
	\includegraphics[width=0.95\columnwidth, trim={0.5cm 0cm 2cm 1.5cm},clip]{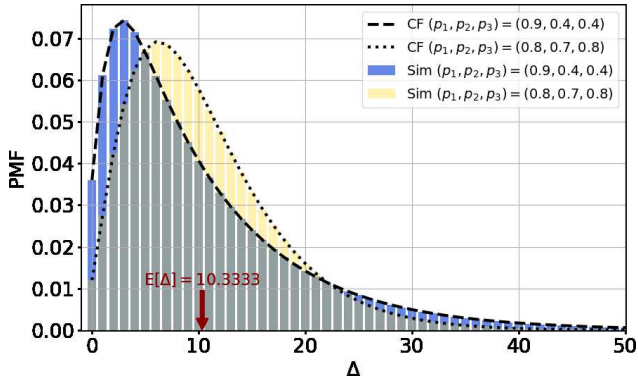}
	\caption{AoI probability mass function of two combinations of three hop loss probabilities $p_l$ with expected age $\mathbb{E}[\Delta_3] = 10.33$.
		Higher loss probability $p_1 = 0.9$ in the first hop increases the distribution tail compared to moderate loss probabilities on all three links.}
	\label{fig:hist}
\end{figure}
\section{Evaluation}
\label{sec:evaluation}

We present a simulation study for the 3-hop case with selected loss probabilities $\left(p_1, \, p_2, \, p_3\right)$ on each link. Each scenario is simulated for $T_{sim} = 100 \, 000$ sampling periods and repeated 100 times. The AoI is evaluated at the end of each sampling period, as described in Sec. \ref{sec:systemmodel}. We measure the expected AoI as:
\begin{equation*}
\E[\Delta_3] \triangleq \dfrac{1}{T_{sim}} \sum_{k=1}^{T_{sim}} \Delta_{3}[k].
\end{equation*}
Note that, one can also use Eq.~\eqref{eq:3hopexpectedAoI} in order to obtain the expected AoI after $3$ hops.
In addition, the \textit{peak age} is defined as the maximum value of AoI at the destination, achieved immediately before receiving a new packet \cite{costa2016on}. We denote the average peak age as $\E[\phi]$.

In order to show the importance of working with a probability distribution instead of expected AoI, we select the following scenarios: $\mathbf{S_1}=\left(p_1=0.9, \,  p_2=0.4, \, p_3=0.4\right)$ and $\mathbf{S_2}=\left(p_1=0.8, \, p_2=0.7, \, p_3=0.8\right)$ which lead to equal expected AoI of $\E\left[\Delta_{3}\right] = 10.33$. $\mathbf{S_1}$ depicts a scenario with two low-loss links and an extremely lossy link, while $\mathbf{S_2}$ represents a scenario where all the links are moderately bad.  Fig. \ref{fig:hist} plots the PMF over AoI up to 50 periods both for the closed form (CF) equation from Eq. \eqref{eq:cf3hop} and for the simulation (Sim). 
We observe that the tail of the PMF is higher for $\mathbf{S_1}$ in comparison to $\mathbf{S_2}$.

\begin{figure}[t]
	\centering
	\includegraphics[width=0.95\columnwidth, trim={0.4cm 0cm 2cm 1.5cm},clip]{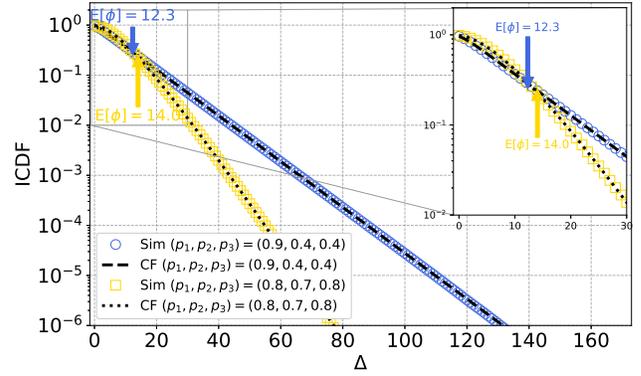}
	\caption{Inverse cumulative distribution function for two combinations of three hop loss probabilities $p_l$ with expected age $\mathbb{E}[\Delta_3] = 10.33$.
		Higher hop probabilities increase the AoI for higher reliability targets. Moreover, for higher reliability targets, the average peak AoI varies drastically compared to the actual AoI.}
	\label{fig:cdf}
\end{figure}

To gain insights into reliability guarantees, we present the inverse cumulative distribution functions (ICDF) in Fig~\ref{fig:cdf}. Despite their equal expected AoI, one can observe that both scenarios pose significant difference beyond $10^{-1}$. In fact, if we are dealing with applications that require high reliability, e.g., five nines, or equivalently $99.999~\%$, Fig. \ref{fig:cdf} shows that both scenarios differ around 20 AoI levels in maximum age. Thus, we conclude that a high loss probability $p_i$ at one of the hops can be fatal for high reliability guarantees. Moreover, it is important to mention that although the $\mathbf{S_1}$ leads to  worst performance in reliability, it achieves lower average peak age than $\mathbf{S_2}$ as shown with arrows in the figure. This leads us to the conclusion that neither the average nor the peak AoI is a sufficient indicator if we want to support applications with reliability guarantees. We need to take the whole probability distribution into account instead. 

%

\section{Conclusion and Future Work}
\label{sec:conclusion}

In this work, we analyze the age of information in a lossy multi-hop network. Lossy multi-hop networks are typical for vehicle-to-vehicle or other machine-to-machine communications. The age of information affects the performance of a set of applications in such scenarios. In the context of AoI, previous work has focused on the expected AoI, the average peak AoI or the maximum AoI. In this work we show that these parameters maybe misleading in a multi-hop network with heterogeneous link loss probabilities. To overcome this problem we provide closed-form expression of the PMF of AoI. 

We believe the closed-form expression of the PMF provides a deeper insight for the age and is useful when it comes to system design choices for high reliability, e.g., five nines, safety-critical communication scenarios such as autonomous cars and automated UAVs. Future work can extend this work to support multiple applications over the same network by adding the queuing perspective at each relay.

\section*{ACKNOWLEDGMENT}
The authors would like to thank the reviewers, whose useful and constructive criticism significantly improved the paper.

\appendix
\subsection{Proof of probability mass functions for different number of hops}
\label{sec:appendic}
In this appendix we share the derivation of probability functions and the expectations.  For instance for the probability of occurrence of any age with 2 hops is
\begin{align}
\label{eq:appendix2hop}
\text{Pr}[\Delta_2[k] = \delta_2]= & \sum_{\delta_1=0}^{\delta_2} (1-p_1) \cdot {p_1}^{\delta_1} \cdot  (1-p_2) \cdot {p_2}^{\delta_2-\delta_1} \nonumber\\
= & (1-p_1)(1-p_2){p_2}^{\delta_2} \cdot \frac{1-\left({\frac{p_1}{p_2}}\right) ^{\delta_2+1}}{1-{\frac{p_1}{p_2}}}\nonumber \\ 
= & (1-p_1)(1-p_2)\cdot \frac{{p_2}^{\delta_2+1}-{p_1}^{\delta_2+1}}{{p_2}-{p_1}}.
\end{align}
For the 3-hop scenario, probability of an age $\delta_3$ can be obtained from Eq.~\eqref{eq:n_hop_age_1} by plugging in our results for $\Delta_2$:
\begin{align*}
\text{Pr}& [\Delta_3[k]=\delta_{3}] = \sum_{\delta_{2}=0}^{\delta_{3}}  (1-p_3){p_3}^{\delta_3-\delta_2} \nonumber \\ &\cdot (1-p_1)(1-p_2)\cdot \frac{{p_2}^{\delta_2+1}-{p_1}^{\delta_2+1}}{{p_2}-{p_1}}\nonumber \\
&= \frac{{p_3}^{\delta_3+1}\prod^3_{i=1} (1-p_i)}{p_2-p_1} \nonumber \\ & \cdot \left( \left(\dfrac{p_2}{p_3}\right)\frac{1-\left({\frac{p_2}{p_3}}\right) ^{\delta_3+1}}{1-{\frac{p_2}{p_3}}} -\left( \dfrac{p_1}{p_3} \right)\frac{1-\left({\frac{p_1}{p_3}}\right) ^{\delta_3+1}}{1-{\frac{p_1}{p_3}}} \right)\nonumber \\
&  = \frac{\prod^3_{i=1} (1-p_i)}{p_2-p_1} \cdot \sum_{j=1}^2 (-1)^j \cdot {p_j} \cdot \frac{{p_3}^{\delta_3+1}-{p_j}^{\delta_3+1}}{{p_3}-{p_j}} 
\end{align*}

\bibliography{eaoi}
\bibliographystyle{IEEEtran}
\end{document}